\documentstyle[twocolumn,aps]{revtex}

  \newcommand{\beq}{\begin{equation}}
  \newcommand{\eeq}{\end{equation}}
  \newcommand{\beql}[1]{\begin{equation}\label{eq:#1}}
  \newcommand{\beqa}{\begin{eqnarray}}
  \newcommand{\eeqa}{\end{eqnarray}}
  \newcommand{\beqas}{\begin{eqnarray*}}
  \newcommand{\eeqas}{\end{eqnarray*}}
  \newcommand{\bA}{{\bf A}}
  \newcommand{\bP}{{\bf P}}
  \newcommand{\bS}{{\bf S}}
  
  \newcommand{\cH}{{\cal H}}
  \newcommand{\cK}{{\cal K}}
  \newcommand{\ket}[1]{|#1\rangle} 
  
  \newcommand{\bracket}[1]{\langle#1\rangle}
  \newcommand{\bx}{{\bf x}}
  
  \newcommand{\da}{\dagger}
  
  \newcommand{\ep}{\epsilon}
  \newcommand{\et}{\eta}
  \newcommand{\eq}[1]{(\ref{eq:#1})} 
  \newcommand{\ga}{\gamma}      
  
  \newcommand{\nn}{\nonumber}

  \newcommand{\ps}{\psi}
  
  \newcommand{\si}{\sigma}             
  \newcommand{\ta}{\tau}   
  \newcommand{\De}{\Delta} 
  \newcommand{\Eq}[1]{Eq.~(\ref{eq:#1})}

\begin{document}
\draft
\title{Position measuring interactions and the Heisenberg
uncertainty principle}
\author{Masanao Ozawa}
\address{Graduate School of Information Sciences, T\^{o}hoku University,
Aoba-ku,  Sendai, 980-8579, Japan}
\maketitle
\begin{abstract}
An indirect measurement model is constructed for an approximately
repeatable, precise position measuring apparatus that violates 
the assertion, sometimes called the Heisenberg uncertainty principle,
that any position measuring apparatus with noise $\ep$ brings the
momentum disturbance no less than $\hbar/2\ep$ in any input state of
the apparatus.
\end{abstract}


Measurements disturb microscopic objects inevitably.  The problem
still remains open as to how measurements disturb their objects.   It is
frequently claimed that if one measures position with noise $\ep$,
the momentum is disturbed at least $\hbar/2\ep$ \cite[p.~230]{vN55}.
This claim is often called the {\em Heisenberg uncertainty principle}.
The Heisenberg principle has been demonstrated typically by a thought
experiment using the $\ga$-ray microscope \cite{Hei81}, and
eventually accepted as a basic principle of quantum mechanics by many
\cite{Jam74}.  However, we have no known general proof for the
Heisenberg principle.

On the other hand, we have another relation claiming that the product of
the standard deviations of position and momentum in any quantum state 
is at least $\hbar/2$.  This relation, often called the {\em
Robertson uncertainty relation}, has been generally proven from two basic
principles of quantum mechanics, the Born statistical formula and the
canonical commutation relation \cite{Ken27,Rob29},
using the Schwarz inequality.   However, this
relation describes the limitation of preparing microscopic objects but has
no direct relevance to the limitation of measurements on single systems
\cite{Bal70,Kra87,MM90}. 

In attempts of formulating the Heisenberg principle in a rigorous
language, there have been serious conceptual confusions concerning the
noise of measurement, as pointed out in Ref.~\cite{88MS} relative to a
controversy as to whether the Heisenberg
principle leads to a sensitivity limit of gravitational wave detection
\cite{Bra74,CTDSZ80,Yue83,Cav85}.
The purpose of this letter is to re-examine the Heisenberg principle by
giving rigorous definitions for noise and disturbance caused by general
measuring interactions.  Two models of position measuring interactions
are examined in detail.  The first one has been known for long
\cite{vN55} and used as a standard model of proposed quantum
nondemolition measurements \cite{CTDSZ80,Cav85}.  By this model
we discuss the justification of our notions of noise and disturbance,
and show how the Heisenberg principle dominates this model.
Then, we modify the first model to obtain the second one which does
not obey the Heisenberg principle.  
From this model, we conclude that  
we have a precise, approximately repeatable position measurement 
that violates the above formulation of the Heisenberg uncertainty
principle.

The disturbance on the object caused by a measurement 
can be attributed to an interaction, called the {\em
measuring interaction}, between the object and the apparatus.  
In this letter, we consider {\em indirect measurement models}
in which the measuring interactions are subject to the equations of
motions in quantum mechanics \cite{00MN,01OD}.

Let $\bA(\bx)$ be a measuring apparatus with macroscopic output
variable $\bx$ to measure, possibly with some error, an observable $A$ of
the {\em object\/} $\bS$, a quantum system represented by a
Hilbert space $\cH$.  The measuring interaction turns on at time $t$, the
{\em time of measurement}, and turns off at time $t+\De t$ between 
object $\bS$ and apparatus $\bA(\bx)$.  
We assume that the object and the apparatus do not 
interact each other before $t$ nor after $t+\De t$ and that 
the composite system $\bS+\bA(\bx)$ is isolated in the
time intervale $(t,t+\De t)$.  The {\em probe} $\bP$ is
defined to be the minimal part of apparatus $\bA(\bx)$ 
such that the composite system $\bS+\bP$ is isolated in the time
intervale $(t,t+\De t)$.
By minimality, we naturally assume that probe $\bP$ is a 
quantum system represented by a Hilbert space $\cK$.  
Denote by $U$ the unitary operator on $\cH\otimes\cK$ representing 
the time evolution of $\bS+\bP$ for the time interval $(t,t+\Delta t)$.  

At the time of measurement the object is supposed to
be in an arbitrary (normalized vector) state $\ps$ and the probe is
supposed to be prepared in a fixed (normalized vector) state $\xi$.
Thus, the composite system $\bS+\bP$ is in the state $\psi\otimes\xi$
at time $t$.  Just after the measuring interaction, the object is separated
from the apparatus, and the probe is subjected to a local interaction with the
subsequent stages of the apparatus.  The last process is assumed to measure
an observable $M$, called the {\em probe observable}, of the probe with
arbitrary precision, and the output is represented by the value of the
macroscopic output variable $\bx$.

In the Heisenberg picture with the original state $\psi\otimes\xi$ at time
$t$, we write 
$A(t)=A\otimes I$,
$M(t)=I\otimes M$,
$A(t+\De t)=U^{\da}(A\otimes I)U$,
and $M(t+\De t)=U^{\da}(I\otimes M)U$.
For any interval $\De$ in the real line, we denote by
``$\bx\in\De$'' the probabilistic event that the output of the
measurement using apparatus $\bA(\bx)$ is in $\De$.
Since the output 
of this measurement is obtained by the measurement of 
the probe observable $M$ at time $t+\De t$,  the probability 
distribution of the output variable $\bx$ is given by
\beql{B1}\label{eq:0328b}
\Pr\{\bx\in\De\}=\bracket{E^{M(t+\De t)}(\De)},
\eeq
where $\bracket{\cdots}$ stands for
$\bracket{\psi\otimes\xi|\cdots|\psi\otimes\xi}$ throughout this letter,
and where $E^{M(t+\De t)}(\De)$ stands for the spectral projection 
of the operator $M(t+\De t)$ corresponding to the interval $\De$.

We say that apparatus $\bA(\bx)$ {\em satisfies the Born statistical
formula (BSF)} for observable $A$ on input state $\ps$, if we have
\beql{010530a}
\Pr\{\bx\in\Delta\}=\bracket{E^{A(t)}(\De)}.
\eeq
We say that apparatus $\bA(\bx)$ {\em measures} observable $A$ {\em
precisely}, if  $\bA(\bx)$ satisfies the BSF for observable $A$ on every
input state \cite{84QC}.  Otherwise, we consider apparatus
$\bA(\bx)$ to measure observable $A$ with some noise.  

In order to quantify the noise, we introduce the {\em noise operator}
$N(A)$ of apparatus $\bA(\bx)$ for measuring $A$ defined by 
\begin{equation}
N(A)=M(t+\Delta t)-A(t). \label{1.1a}
\end{equation}
The {\em noise} $\ep(A)$ of apparatus $\bA(\bx)$ for
measuring $A$ on input state $\ps$ is, then,  defined by 
\beql{noise}
\ep(A)=\bracket{N(A)^{2}}^{1/2}.
\eeq
The noise $\ep(A)$ represents the 
root-mean-square error in the output of the measurement.

In order to clarify the meaning of the above definition, let us consider
the case where the measured observable has a definite value just before
the measurement, so that we assume $\ps=\ket{A=a}$.
Then, we have
\beq
N(A)\ket{\ps\otimes\xi}=[M(t+\Delta t)-a]\ket{\ps\otimes\xi}
\eeq
and 
\beq
\ep(A)=\bracket{[M(t+\Delta t)-a]^{2}}^{1/2}.
\eeq
Thus, $\ep(A)$ stands for the root-mean-square deviation in the experimental
output $M(t+\Delta t)$ from the value $a$ of observable $A$ taken 
at the time of measurement.

One of the fundamental properties of the noise is that precise apparatuses
and noiseless apparatuses are equivalent notions, as ensured by the
following theorem \cite{note1}.

{\bf Theorem 1.}
{\em Apparatus $\bA(\bx)$ measures observable
$A$ precisely if and only if $\ep(A)=0$ on any input state $\ps$.}

In this letter, we assume that the apparatus carries out
instantaneous measurements.  In this case, we say that apparatus
$\bA(\bx)$ {\em does not disturb the probability distribution of}
an observable $B$ of $\bS$
on input state $\ps$, if 
\beql{829e}
\bracket{E^{B(t)}(\De)}=\bracket{E^{B(t+\De t)}(\De)}
\eeq
for every interval  $\De$, where 
we write $B(t)=B\otimes I$ and $B(t+\Delta
t)=U^{\dagger}(B\otimes I)U$.  We say that apparatus $\bA(\bx)$ {\em
does not disturb} observable $B$, if apparatus $\bA(\bx)$ does not disturb
the probability distribution of  observable $B$ on any input state $\ps$
\cite{01OD}. It was proven that apparatus $\bA(\bx)$ does not disturb
observable $B$ if and only if successive measurements of observables $A$
and $B$, using $\bA(\bx)$ for $A$ measurement, satisfy the joint
probability formula for simultaneous measurements \cite{01OD}.

In order to
quantify the disturbance, we introduce the {\em disturbance operator}
$D(B)$ of apparatus $\bA(\bx)$ for observable $B$ defined by 
\beql{disturbance_operator}
D(B)=B(t+\Delta t)-B(t).
\eeq
The {\em disturbance} $\et(B)$ of apparatus $\bA(\bx)$ for
observable $B$ on input state $\ps$ is, then,  defined by 
\beql{disturbance}
\et(B)=\bracket{D(B)^{2}}^{1/2}.
\eeq
The disturbance $\et(B)$ represents the root-mean-square 
deviation of the observable $B$ before and after the measuring
interaction. 

One of the fundamental properties of the disturbance is that
apparatuses that do not disturb (the probability distribution of)
the given observable 
and apparatuses with zero disturbance for that observable 
are equivalent notions, as ensured by the following theorem \cite{note1}.

{\bf Theorem 2.}
{\em Apparatus
$\bA(\bx)$ does not disturb  observable $B$ if and only if $\et(B)=0$ on
any input state $\ps$.}

From now on, we consider the case where the object  $\bS$ is a
one-dimensional mass with position  $\hat{x}$ and  momentum
$\hat{p}_{x}$.
Under general definitions given in the previous sections, 
we can rigorously formulate the Heisenberg uncertainty principle that 
any position measurement with noise $\ep$ disturbs
the momentum at least $\hbar/2\ep$ by the relation
\beql{Heisenberg}
\ep(\hat{x})\et(\hat{p}_{x})\ge\frac{\hbar}{2}.
\eeq

Von Neumann
\cite[p.~443]{vN55} introduced the following indirect measurement
model of an approximate position measurement (see also
Refs.~\cite{CTDSZ80,Cav85,93CA}).  The probe $\bP$ is 
another one-dimensional mass with  position
$\hat{y}$ and momentum $\hat{p}_{y}$.   The probe observable is
taken to be position $\hat{y}$. The measuring interaction is given by 
\beql{829o} 
\hat{H}=K\hat{x}\hat{p}_{y}.
\eeq 
The coupling constant  $K$ is so large that the free Hamiltonians
can be  neglected.  The time duration  $\De t$ of the measuring
interaction is chosen so that $K\De t=1$.  Then, the unitary operator
of the time evolution of $\bS+\bP$ from $t$ to $t+\De t$ is given by
\beql{829p}
U=\exp\left(\frac{-i}{\hbar}\hat{x}\hat{p}_{y}\right).
\eeq

Solving the Heisenberg equations of motion for $t<t+\ta<t+\De t$, we
obtain 
\begin{mathletters}
\beqa
\hat{x}(t+\ta)&=&\hat{x}(t),\\
\hat{y}(t+\ta)&=&K\ta\hat{x}(t)+\hat{y}(t),\\
\hat{p}_{x}(t+\ta)&=&\hat{p}_{x}(t)-K\ta\hat{p}_{y}(t),\\
\hat{p_{y}}(t+\ta)&=&\hat{p}_{y}(t).
\eeqa
\end{mathletters}%
For $\ta=\De t=1/K$, we have
\begin{mathletters}
\beqa
\hat{x}(t+\De t)&=&\hat{x}(t),\\
\hat{y}(t+\De t)&=&\hat{x}(t)+\hat{y}(t),\\
\hat{p}_{x}(t+\De t)&=&\hat{p}_{x}(t)-\hat{p}_{y}(t),\\
\hat{p_{y}}(t+\De t)&=&\hat{p}_{y}(t).
\eeqa
\end{mathletters}%
It follows that the noise operator and the disturbance operator are
given by
\begin{mathletters}
\beqa
N(\hat{x})
&=&\hat{y}(t+\De t)-\hat{x}(t)=\hat{y}(t),\\
D(\hat{p}_{x})
&=&\hat{p}_{x}(t+\De t)-\hat{p}_{x}(t)=-\hat{p}_{y}(t).
\eeqa\end{mathletters}%
Thus, the position-measurement noise and the momentum disturbance
are given by
\begin{mathletters}\beqa
\ep(\hat{x})^{2}
&=&\bracket{\hat{y}(t)^{2}},\\
\et(\hat{p}_{x})^{2}
&=&\bracket{\hat{p}_{y}(t)^{2}}.
\eeqa\end{mathletters}%
We denote by $\si(\hat{y})$ and $\si(\hat{p}_{y})$ the standard
deviations of the probe position and momentum at
the time of measurement, respectively.  
By definition, we have
\begin{mathletters}\beqa
\si(\hat{x})^{2}
&=&\bracket{\hat{y}(t)^{2}}-\bracket{\hat{y}(t)}^{2}
\le\ep(\hat{x})^{2},\\
\si(\hat{p}_{x})^{2}
&=&\bracket{\hat{p}_{y}(t)^{2}}-\bracket{\hat{p}_{y}(t)}^{2}
\le\et(\hat{p}_{x})^{2}.
\eeqa\end{mathletters}%
Thus, by the Robertson relation, we have
\beq
\ep(\hat{x})\et(\hat{p}_{x})\ge\si(\hat{y})\si(\hat{p}_{y})
\ge\frac{\hbar}{2}.
\eeq

Therefore, we conclude that the von Neumann model obeys 
\Eq{Heisenberg} as a consequence of the
Robertson relation  applied to the probe state just before
measurement.  In particular, this model represents a basic feature of the
$\ga$ ray  microscope on the point that the trade-off between the noise
and the disturbance arises from the fundamental physical limitation on
preparing the probe. It might be expected that such a basic feature is shared by
every model in a reasonable class of position measurements.  However,
the next model suggests that it is not the case.

In what follows, we modify the measuring interaction of the von 
Neumann model to construct a model that violates \Eq{Heisenberg}. 
In this new model, the object, the probe, and the probe
observables are the same systems and the same observable as the von
Neumann model. The measuring interaction is taken to be \cite{88MS}
\beql{829ox}
\hat{H}=\frac{K\pi}{3\sqrt{3}}
(2\hat{x}\hat{p}_{y}-2\hat{p}_{x}\hat{y}
+\hat{x}\hat{p}_{x}-\hat{y}\hat{p}_{y}).
\eeq
The coupling constant  $K$ and the time duration  $\De t$ are chosen as
before so that $K\gg 1$ and $K\De t=1$.
Then, the unitary operator $U$ is given by 
\beql{829px}
U=\exp\left[\frac{-i\pi}{3\sqrt{3}\hbar}
(2\hat{x}\hat{p}_{y}-2\hat{p}_{x}\hat{y}
+\hat{x}\hat{p}_{x}-\hat{y}\hat{p}_{y})\right].
\eeq

Solving the Heisenberg equations of motion for $t<t+\ta<t+\De t$, we
obtain 
 \begin{mathletters}\beqa
 \lefteqn{\hat{x}(t+\ta)}\quad\nn\\
 &=&\frac{2}{\sqrt{3}}\hat{x}(t)\sin \frac{(1+K\ta)\pi}{3}
 +\frac{-2}{\sqrt{3}}\hat{y}(t)\sin \frac{K\ta\pi}{3},\\
 \lefteqn{\hat{y}(t+\ta)}\quad\nn\\
 &=&\frac{2}{\sqrt{3}}\hat{x}(t)\sin \frac{K\ta\pi}{3}
 +\frac{-2}{\sqrt{3}}\hat{y}(t)\sin \frac{(1-K\ta)\pi}{3},\\
 \lefteqn{\hat{p}_{x}(t+\ta)}\quad\nn\\
 &=&\frac{-2}{\sqrt{3}}\hat{p}_{x}(t)\sin \frac{(1-K\ta)\pi}{3}
 +\frac{-2}{\sqrt{3}}\hat{p}_{y}(t)\sin \frac{K\ta\pi}{3},\\
 \lefteqn{\hat{p}_{y}(t+\ta)}\quad\nn\\
 &=&\frac{2}{\sqrt{3}}\hat{p}_{x}(t)\sin \frac{K\ta\pi}{3}
 +\frac{2}{\sqrt{3}}\hat{p}_{y}(t)\sin \frac{(1+K\ta)\pi}{3}.
 \eeqa\end{mathletters}%
For $\ta=\De t=1/K$, we have
\begin{mathletters}\beqa
\hat{x}(t+\De t)&=&\hat{x}(t)-\hat{y}(t),\label{eq:ozawa-model-1}\\
\hat{y}(t+\De t)&=&\hat{x}(t),\label{eq:ozawa-model-2}\\
\hat{p}_{x}(t+\De t)&=&-\hat{p}_{y}(t),\label{eq:ozawa-model-3}\\
\hat{p}_{y}(t+\De t)
&=&\hat{p}_{x}(t)+\hat{p}_{y}(t)\label{eq:ozawa-model-4}.
\eeqa\end{mathletters}%
It follows that the noise operator and the disturbance
operator are given by
\begin{mathletters}\beqa
N(\hat{x})
&=&\hat{y}(t+\De t)-\hat{x}(t)=0,\\
D(\hat{p}_{x})
&=&\hat{p}_{x}(t+\De t)-\hat{p}_{x}(t)=-\hat{p}_{y}(t)
-\hat{p}_{x}(t).
\eeqa\end{mathletters}%
Thus, the position-measurement noise and the
momentum disturbance are given by
\begin{mathletters}\beqa
\ep(\hat{x})
&=&0,\\
\et(\hat{p}_{x})^{2}
&=&\bracket{[\hat{p}_{x}(t)+\hat{p}_{y}(t)]^{2}}\nn\\
&=&\si(\hat{p}_{x})^{2}+\si(\hat{p}_{y})^{2}
+[\bracket{\hat{p}_{x}(t)}+\bracket{\hat{p}_{y}(t)}]^{2}.\nn\\
\eeqa\end{mathletters}%
Consequently,  we have \cite{note2}
\beq
\ep(\hat{x})\et(\hat{p}_{x})=0.
\eeq
Therefore, our model obviously violates \Eq{Heisenberg}.  

Taking advantage of 
the above model, we can refute the argument
that the uncertainty principle generally leads to a general sensitivity 
limit, called the standard quantum limit, for monitoring free-mass
position \cite{Yue83,88MS}. 

If $\bracket{\hat{p}_{x}(t)^{2}}\to 0$
and $\bracket{\hat{p}_{y}(t)^{2}}\to 0$ (i.e.,  $\ps$ and $\xi$
tend to the momentum eigenstate with zero momentum) then
we have even $\et(\hat{p}_{x}(t))\to 0$ with $\ep(\hat{x})=0$.
Thus, we can measure position precisely without effectively disturbing 
momentum in a near momentum eigenstate;
see Ref.~\cite{01CQSR} for detailed discussion on the 
quantum state reduction caused by the above model.

In formulating the canonical description of state changes caused by
measurements,
von Neumann required not only the preciseness of measurement but
also that the measurement of an observable satisfy
the repeatability hypothesis  \cite[p.~335]{vN55}: {\em If an observable is
measured twice in succession in a system, then we get the same value
each time.}
On the other hand, the von Neumann model \eq{829p}
does not satisfy the preciseness
nor the repeatability.
One of the characteristic features of our model \eq{829px}
is that it measures position precisely, but our model does not satisfy the
repeatability hypothesis either.  
Thus, it is tempting to understand that our model circumvents the
Heisenberg inequality \eq{Heisenberg} by paying the price of failing the
repeatability.  Nevertheless, the following argument will show
that such a view cannot be supported.  

In the first place, it has been proven
that the repeatability hypothesis can be satisfied only by measurements   
of purely discrete observables.  Davies and Lewis \cite{DL70} gave a 
mathematical formulation of the repeatability hypothesis in a form that
is meaningful even for measurements of continuous observables,
and yet conjectured that it can be satisfied only by 
measurements of purely discrete observable.  This conjecture was
actually proven affirmatively \cite{84QC,85CA}.
Thus, no precise position measurements satisfy the repeatability hypothesis.

Secondly, if we consider the approximate repeatability, our model
satisfies any stringent requirement on approximate repeatability.
In order to show this, we need to introduce the measure 
of approximate repeatability.  Suppose that we measure the position 
of mass $\hat{x}$ in succession using two apparatuses described 
by equivalent indirect measurement models.   
Suppose that the first apparatus with probe $\hat{y}$ interacts with
$\hat{x}$ in $(t,t+\Delta t)$ 
and that the second apparatus with probe $\hat{z}$
interacts with $\hat{x}$ in $(t+\Delta t, t+2\Delta t)$.
Then, for any positive number $\alpha$, 
the position measurement is called an {\em $\alpha$ approximately
repeatable}, if the root-mean-square deviation between the first output 
$\hat{y}(t+\De t)$ 
and the second output $\hat{z}(t+2\De t)$ is no more than $\alpha$
\cite{93CA}, i.e., 
\beq
\bracket{\ps\otimes\xi\otimes\xi|
[\hat{z}(t+2\De t)-\hat{y}(t+\De)]^{2}|\ps\otimes\xi\otimes\xi}\quad\\
\le\alpha^{2}.
\eeq
If the apparatuses are equivalent to our model \eq{829px}, 
we have
\beq
\hat{z}(t+2\De t)=\hat{x}(t+\De t)=\hat{x}(t)-\hat{y}(t),
\eeq
and hence
\beqas
\lefteqn{
\bracket{\ps\otimes\xi\otimes\xi|
[\hat{z}(t+2\De t)
-\hat{y}(t+\De t)]^{2}|\ps\otimes\xi\otimes\xi}}\qquad
\\ &=&
\bracket{\hat{y}(t)^{2}}.\qquad\qquad\qquad\qquad\qquad\qquad\qquad
\eeqas
Thus, assuming $\bracket{\hat{y}(t)}=0$, we can conclude that our model
is $\si(\hat{y})$ approximately repeatable.  

Therefore, we conclude that although we have no repeatable position 
measurements in general,
for any small $\alpha>0$ we have a precise, $\alpha$ repeatable
position measurement that violates the Heisenberg inequality
\eq{Heisenberg}.
This suggests that how stringent conditions might be
posed for a class of position measurements, 
we could find at least one model
that violates the Heisenberg inequality \eq{Heisenberg} in that class.

In their discussion on the Heisenberg principle, 
Braginsky and Khalili \cite[p.~65]{BK92} claimed that if the object input
state is near a momentum eigenstate, then the post-measurement position
uncertainty, $\si(\hat{x})(t+\De t)$, will be equal to the noise, 
$\ep(\hat{x})$.  This claim and the subsequent derivation of the Heisenberg
inequality is incorrect, since our model
shows that $\si(\hat{x})(t+\De t)\to\infty$ and $\ep(\hat{x})=0$
when $\ps$ goes to the momentum eigenstate $\ket{\hat{p}_{x}=0}$,
so that they can never be close.

Braginsky and Khalili \cite[p.~66]{BK92} claimed also that all linear
measurements,  measurements closely connected to linear systems, obey
the Heisenberg inequality.  
However, our unitary operator \eq{829px} can be
realized by linear systems as follows.  The Hamiltonian
\eq{829ox} of our model comprises simple linear couplings
$\hat{x}\hat{p}_{y}$ and 
$\hat{p}_{x}\hat{y}$ and an extra term
$\hat{x}\hat{p}_{x}-\hat{y}\hat{p}_{y}$, which might resist
a simple linear realization.  However, the extra term can be eliminated
by the following mathematical relation \cite{01CQSR}
\beqa
\lefteqn{\exp\left[\frac{-i\pi}{3\sqrt{3}\hbar}
(2\hat{x}\hat{p}_{y}-2\hat{p}_{x}\hat{y}
+\hat{x}\hat{p}_{x}-\hat{y}\hat{p}_{y})\right]}\quad\quad\nn\\
&=&\exp\left(-\frac{i}{\hbar}\hat{x}\hat{p}_{y}\right)
  \exp\left(\frac{i}{\hbar}\hat{p}_{x}\hat{y}\right).
\label{eq:realization}
\eeqa
Thus, our measuring interaction, \eq{829px},  is equivalent to
the consecutive linear couplings $\hat{p}_{x}\hat{y}$ and
$\hat{x}\hat{p}_{y}$ \cite{01CQSR}.
The interactions corresponding to $\hat{p}_{x}\hat{y}$ and
$\hat{x}\hat{p}_{y}$ have been known as the back-action evading
(BAE) measurement \cite{Yur85} and its conjugate.  They have
been experimentally realized in linear optics \cite{POK94,BLGL95}
with equivalent optical setting of quadrature measurements.  
According to the above, our model can be experimentally
realized at least in an equivalent optical setting using current linear
optical devices, a combination of two mutually conjugate
back-action evading amplifiers.

In Ref.~\cite{01OD}, it was proven that any measuring 
apparatus disturbs every observable not commuting 
with the measured observable.
Our model \eq{829px} suggests, however,  that 
the trade-off between noise and disturbance 
should be quantitatively represented by a more complex formula 
than the Heisenberg inequality \eq{Heisenberg}.
Instead of \Eq{Heisenberg}, our model \eq{829px} actually 
satisfies the following trade-off between
the initial position uncertainty $\si(\hat{x})$ and the momentum
disturbance $\et(\hat{p}_{x})$ as
\beql{noiseless}
\si(\hat{x}) \et(\hat{p}_{x})\ge \frac{\hbar}{2}.
\eeq
A proof of the above relation runs as follows.
From \Eq{ozawa-model-3}, we have
\beq
[\hat{x}(t),D(\hat{p}_{x})]=[\hat{x}(t),-\hat{p}_{x}(t)]=-i\hbar.
\eeq
Thus, by the Robertson relation, we have
\beq
\si(\hat{x})\, \et(\hat{p}_{x})
\ge\frac{1}{2}|\bracket{[\hat{x}(t),D(\hat{p}_{x})]}|
=\frac{\hbar}{2}.
\eeq
A universally valid trade-off relation for the noise in position, 
the disturbance in momentum, and the initial uncertainties 
of position and momentum extending relations \eq{Heisenberg}
and \eq{noiseless} to all generalized measurements
will be shown in a forthcoming paper.

{\bf Acknowledgments.} 
This work was supported by the programme ``R\&D on Quantum  
Tech.'' of the MPHPT of Japan, by the CREST project of the JST,
and by the Grant-in-Aid for Scientific Research of the JSPS.


\begin{thebibliography}{00}
\bibitem{note1} A proof will be given in a forthcoming paper.
\bibitem{note2} Mathematically, we assume here that state $\ps$ is 
in a common domain of $\hat{x}$ and
$\hat{p}_{x}$, while state $\xi$ is in a common domain of $\hat{y}$ 
and $\hat{p}_{y}$.
\end{thebibliography}
\end{document}